\documentstyle[12pt,epsfig]{article}
\begin{document}
\begin{titlepage}

\begin{flushright}
  \bf{CMS NOTE 2001/010}
\end{flushright}
  
\begin{center}
  {\Large \bf Non-linear energy calibration of CMS calorimeters for single pions}
\end{center}

\vspace{10mm}

\begin{center}
    J. Damgov, V. Genchev, S. Cht. Mavrodiev

       INRNE, Sofia, Bulgaria
\end{center}

\begin{center}
(February 26, 2001)
\end{center}

\vspace{10mm}

\begin{abstract}
 CMS calorimeter energy calibration was done in the full CMS simulated 
 geometry for the pseudorapidity region $\eta$ = 0. The samples of
 single pion events were generated with a set of incident energies from 10 
 GeV to 3 TeV. The analysis of the simulated data shows 
 that standard calibration using just sampling coefficients for calorimeter 
 parts with different sampling ratio gives nonlinear calorimeter 
 response. Non-linear calibration technique was applied for improving 
 calorimeter energy resolution and restoring the calorimeter linearity.
\end{abstract} 

\vspace{80mm}

This study is supported by the Bulgarian Ministry of Education and Sciences
  
\end{titlepage}

\setcounter{page}{2}

\section{Calorimeters geometry considered}

 The pion responses are obtained with GEANT simulations with a 
detailed description of CMS calorimeter geometry (CMSIM 115), 
version TDR-2$^{/1/}$.

The ECAL is the PbWO$_{4}$ one (readout number 1).

An additional 1 cm scintillator layer is placed in front of HB
to compensate for the energy loss in the cables, electronics and cooling 
system on the back side of the ECAL, which is simulated by a uniform slab 
of 0.2 $\lambda$ of dead material (readout number 2).
 
The sampling thicknesses of copper alloy (90\%Cu + 10\%Zn) are 5 cm in the 
barrel segments, except the inner and outer plates 
which are 7 cm stainless steel. The 9 mm gaps between the absorber plates 
are filed with 4 mm scintillator planes (readout number 3). 

To improve the energy measurement for late developing hadron showers, 
tail catcher layers are inserted in front and behind the first return 
yoke iron layer (RY1) (readout number 4).  
 
The regions behind the ECAL are empty except for 
the scintillator layer and equivalent dead material uniform absorber layer. 
The effects of the CMS 4T field are fully included. 

\section{Data sample}

We have performed a GEANT simulation of the response to pions for a set 
of incident energies from 10 to 3000 GeV at pseudorapidity set to 0. GHEISHA
was used as a hadron interaction simulator.
The sum of the energies deposited by the showers in the active elements of
each readout has been stored 
in a disk file so that the showers could be analysed later. 

\section{Standard calibration}

The standard calibration used just sampling coefficients for calorimeter 
parts with different sampling ratio. 
The reconstructed energy of simple shower is given by the weighted sum
of the energies deposited in the readouts:

 $$ E^{rec} = \sum_{i=1, 4} C_{i} E_{i} , $$

where:

$E_{i}$ - amplitude of the signal from the calorimeter longitudinal 
segmentation (readouts);

$C_{i}$ - calibration coefficients.

Coefficients $C_{i}$, i = 1, 4 are determined by the minimisation of the width 
of the energy distributions.

The Gaussian part of reconstructed energy distributions at various incident 
energies are then fitted to obtain the calorimeter energy 
resolution. The energy resolution is parametrised by the expression:

$$ \sigma /E = a/ \sqrt{E} \oplus b. $$ 

The obtained energy resolution by the standard calibration is shown on fig. 1 
and the residuals of the reconstructed energy on fig. 2.    

\section{Non-linear technique}

The nonlinear behaviour of the calorimeter response could be solved by 
application of the non-linear method which improve 
the linearity of the calorimeter response and the energy resolution in the 
broad energy range.
Non-linear technique is the selection of some additional parameters which 
provide correct energy reconstruction of the hadron shower.

Thus the reconstructed energy $E^{rec}_{nl}$ is parametrised as:

 $$ E^{rec}_{nl} = \sum_{i=1, 4} f_{i}(\vec{A}, E_{i}) E_{i} , $$

where $f_{i}$ are non-linear functions of unknown parameters $\vec{A}$.

Large number of free parameters (more than 50) are needed to better
describe the non-linearity. For such a big number of parameters
minimisation performed by MINUIT is inefficient. To this end we used
the specialised program REGN based on a autoregularized Newton type
method$^{/2/}$. In the REGN computer code$^{/3/}$ the $\chi^{2}$ is one
of the criteria available for solving the system and for testing the
mathematical model. Other criteria permit one to chose uniquely between
several model functions the best one$^{/4, 5, 6/}$.

The energy resolution obtained  applying non-linear calibration
method is shown in fig. 1 together with results from
the standard calibration.
In Figure 2 the residuals of the reconstructed energies
are shown for both methods.
The comparison of the standard
calibration method to the non-linear technique shows clear
improvement in the resolution and  linearity of the reconstructed energy.

\section{Summary}

The CMS calorimeters pion calibration was done using two different approaches.
Calibration using non-linear technique improves the resolution and  linearity
of the reconstructed energy.


\vspace{30mm}

\begin{figure}[hbtp]
  \begin{center}
    \resizebox{14cm}{!}{\includegraphics{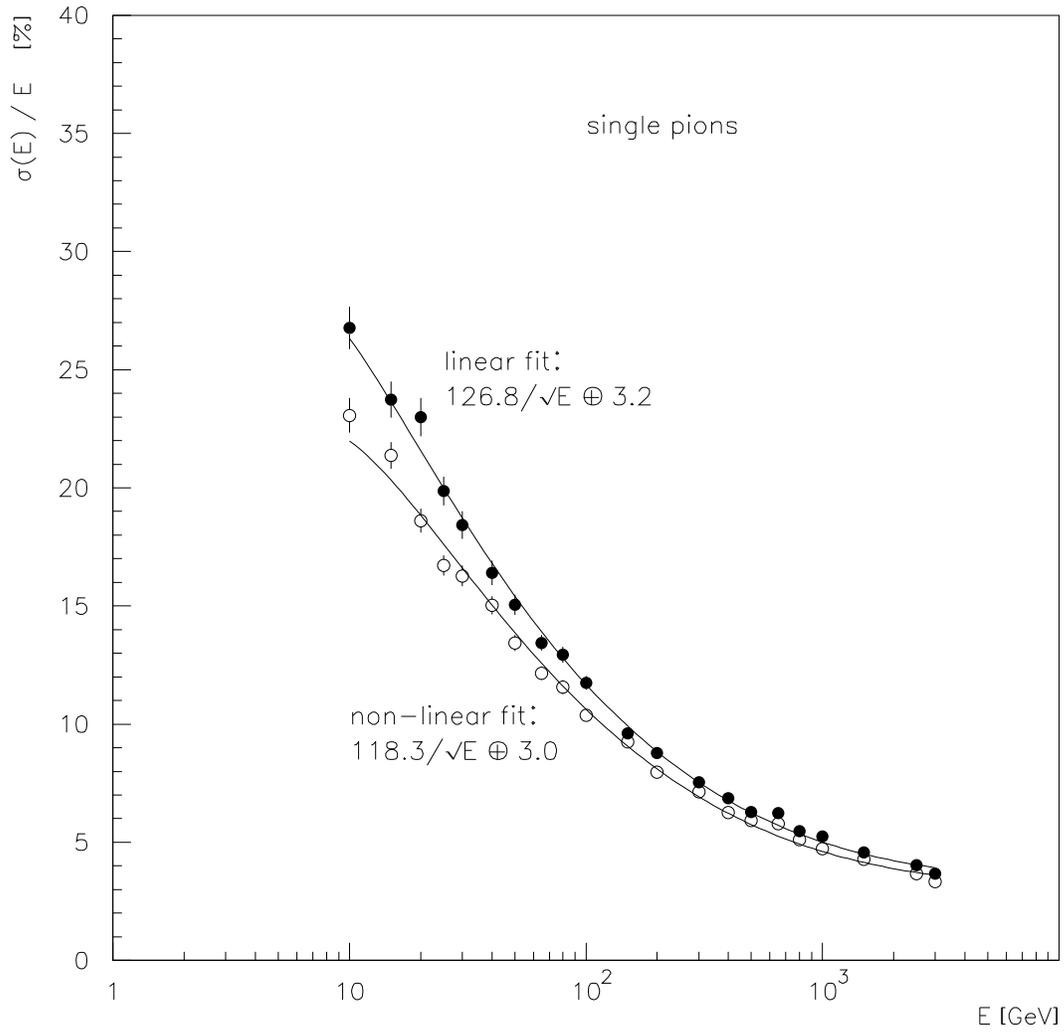}}
\caption{Energy resolution for the CMS calorimeter system (standard
calibration - full points, non-linear calibration - open points).} 
\label{fig:1}
  \end{center}
\end{figure}

\begin{figure}[hbtp]
  \begin{center}
    \resizebox{14cm}{!}{\includegraphics{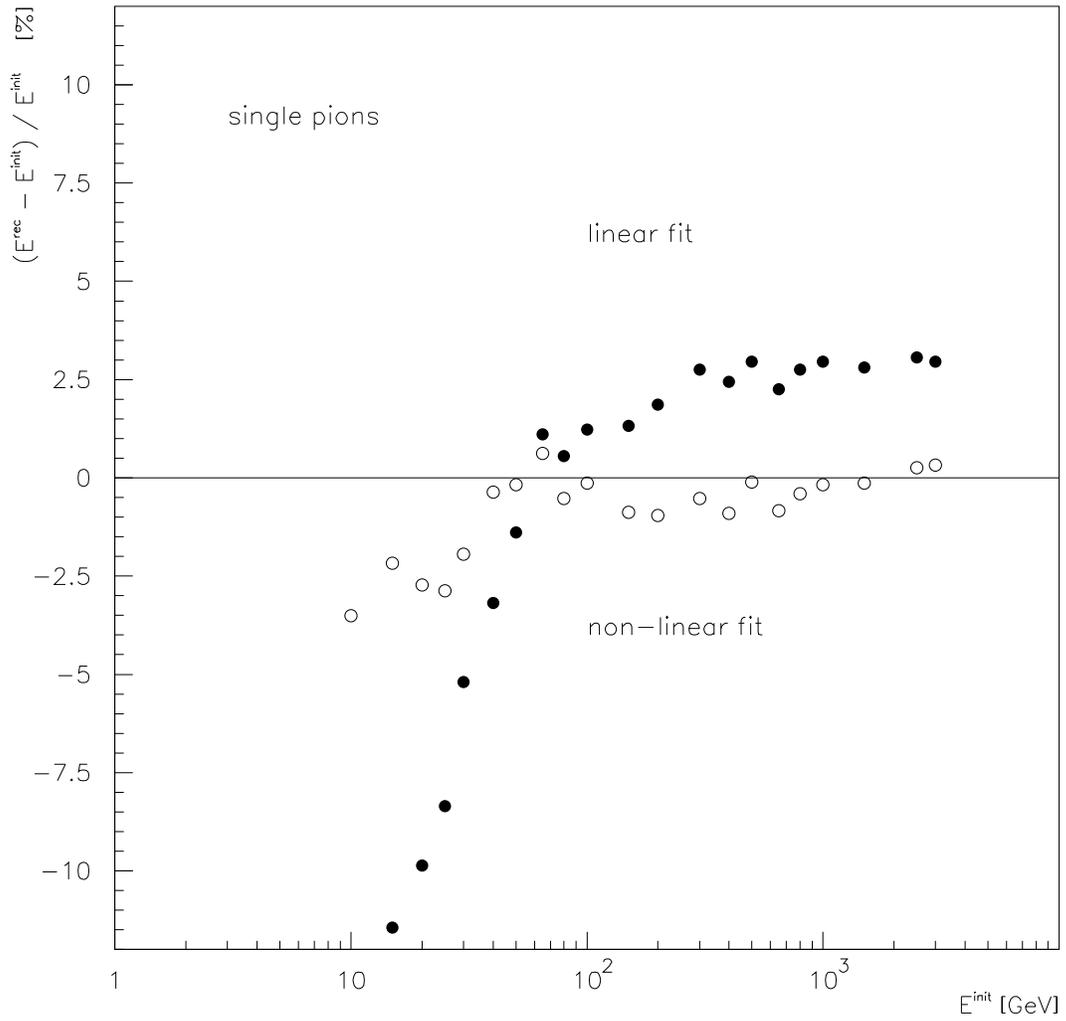}}
\caption{Residuals of the reconstructed energy (standard
calibration - full points, non-linear calibration - open points).} 
\label{fig:2}
  \end{center}
\end{figure}

\end{document}